\documentstyle[12pt,epsfig]{article}
\font\petit = cmr10
\begin{document}
\title{
First and second order transition of frustrated Heisenberg spin systems
	}
\author{D. Loison$^{1,2}$ and K.D. Schotte$^1$
\thanks{loison@physik.fu-berlin.de, schotte@physik.fu-berlin.de}
\\
$^1${\petit Institut f\"ur Theoretische Physik, Freie Universit\"at Berlin}\\
{\petit Arnimallee 14, 14195 Berlin, Germany}\\
{\petit and}\\
$^2${\petit Laboratoire de Physique Th\'eorique et Mod\'elisation, 
Universit\'{e} de Cergy-Pontoise}\\
{\petit 2, Av.\thinspace A.~Chauvin, 95302 Cergy-Pontoise Cedex, France}
	}
\maketitle
\begin{abstract}
Starting from the hypothesis of a second order transition we have
studied modifications of the
original Heisenberg antiferromagnet on a stacked triangular
lattice (STA--model) by the Monte Carlo technique. 
The change is a local constraint restricting
the spins at the corners of selected triangles to add up to zero without
stopping them from moving freely (STAR--model). 
We have studied also the closely
related dihedral and trihedral models which can be classified as Stiefel
models. We have found indications of a first order transition for all three 
modified models instead of a universal critical behavior. 
This is in accordance with the renormalization group investigations but
disagrees with the Monte Carlo simulations of the original STA--model
favoring a new universality class. For the corresponding
$x$--$y$ antiferromagnet studied before, the second order
nature of the transition could also not be confirmed.

\end{abstract}
\vspace{1.cm}
P.A.C.S. numbers:05.70.Fh, 64.60.Cn, 75.10.-b 
\section{INTRODUCTION}

The critical properties of frustrated spin systems are still under
discussion \cite{Diep2}. In particular no consensus exists about the
nature of the phase transition of an antiferromagnet on a stacked
triangular lattice with nearest neighbor interactions. The
symmetry group for a Heisenberg antiferromagnet at high temperatures $SO_3$
changes to $SO_2$ for the usual antiferromagnet, but the symmetry is
completely broken in the low temperature phase of the frustrated
antiferromagnet. This difference in the breakdown of symmetry between
the frustrated and non frustrated cases should lead to
different classes of critical behavior. The controversial point
for the stacked triangular antiferromagnet (STA) is whether the
phase transition is of second order with a new chiral universality
class \cite{Kawa88} or whether its true nature is a weak first order change.

Monte Carlo studies 
\cite{Kawa92,PlumerHei,LoisonHei,Bhattacharya,Diep89,Loisonbct} 
favor a new universality class whereas
renormalization group studies indicate a first order transition,
since no stable fixed point can be found for the Heisenberg case
in order $\epsilon^2$ with $\epsilon = 4 - d$ (d is the dimension
of space) \cite{Antonenko2} and also for an expansion up to three loops in
$d = 3$ \cite{Antonenko 94}. 
The results for the critical exponents of the numerous
Monte Carlo simulations are listed in table \ref{table1}. 
They agree reasonably well and also with the experimental values 
in table \ref{table3}.
One notices that the new critical indices are quite distinct from
the standard ones for the Heisenberg model
with symmetry breaking $SO_3/SO_2$ listed in table \ref{table2}. 
The only flaw is the small negative values of $\eta$, since this exponent 
should always be positive \cite{Patashinskii,Zinn89}.

The discrepancy between the renormalization group
analysis and the Monte Carlo simulation can be removed by assuming that
the first order phase transition occurs with a correlation length larger
than the diameter of the cluster studied. Since this size cannot be
substantially increased with the present technical means and skills
we have tried to investigate modified lattice spin systems which should 
belong to the same universality class. The idea is to shorten the correlation 
length by these modifications in order to reveal the
first order nature of the transition. This strategy was successful
for the $x$--$y$ STA--model we investigated before \cite{LoisonSchotteXY} 
and we could show that in fact the phase transition is of first order 
in this case.

Following Zumbach \cite{Zumbach93} 
one can analyze such a weak first order transition
in terms of the renormalization group approach as due to a fixed point
in the complex parameter space 
(More precisely it is only necessary to have a minimum in the RG flow). 
A basin of attraction for the real
parameters will be generated and ``mimic'' a second order transition
with slightly changed scaling relations. The unphysical negative values
of $\eta$ in table \ref{table1} could be corrected using Zumbach's approach.

Investigations about the phase transitions of the STA--systems
and similar helimagnetic systems \cite{Diep89,Loisonbct} are not only of 
interest for the field of magnetism including the experimental studies.
Phase transitions in superfluids, in type II superconductors,
and in smectic--A liquid crystals should be similar in nature to the ones
of frustrated Heisenberg magnets.

The numerical simulations are closely connected to a similar study for
the $x$--$y$ spins \cite{LoisonSchotteXY}. 
We study the classical Heisenberg spin system
on the stacked triangular lattice by fixing for selected triangles the spin
direction to a 120$^{\circ}$--order which would correspond to the ideal
antiferromagnetic order on a triangular lattice as in Fig.~\ref{figure24}.
The common orientation of a 120$^{\circ}$--cluster is still free.
The principle behind this construction is that modes removed by the
120$^{\circ}$--rigidity are ``irrelevant'' close to the critical temperature.
What is relevant is the common direction and orientation. The antiferromagnet
STA and the rigid antiferromagnet STAR should therefore be in the same
universality class. These considerations can also be applied to Stiefel's
$V_{3,2}$ or dihedral model \cite{Kunz}. 
A third model studied is the right handed
trihedral model. It is constructed by adding a third vector given by the vector
product of the two of the dihedron. As we do not add new degrees of freedom the
model should again belong to the same universality class.

That the behavior close to the phase transition should not change
if one modifies the models by constraints can only be expected
from a second order transition. However, if one assumes that the
transition is of first order, but not visible since the barriers
between the two phases are too low for the spin clusters one can
simulate, then building in constraints in the models and simulations should
make a difference since the barriers are more difficult to overcome.

In section II we present a review of the RG studies. 
In the following section the three models studied are presented.
The details of the simulations and the finite size scaling analysis
are described in section IV. Results are given in section V. 
The discussion is in section VI where we will review also
the experimental studies and the other RG studies in $2+\epsilon$ 
expansions.

\section {Renormalization group studies and complex fixed points \label{RGZ}}
\subsection{Renormalization group studies in $d=3$ and $d=4-\epsilon$}

The renormalization group studies have motivated
for the re-analysis of the frustrated Heisenberg antiferromagnet with
the Monte Carlo technique. Therefore we repeat briefly the main points.
For more details we refer to Antonenko and Sokolov \cite{Antonenko 94}.

The Hamiltonian one considers is  
\begin{equation}
\label{Hamiltonian}
H={1 \over 2} \int d^3x \Bigl{[}r_0^2 \,\phi_\alpha \phi_\alpha^*+
\nabla \phi_\alpha \nabla\phi_\alpha^* + 
{u_0 \over 2} \phi_\alpha \phi_\alpha^* \phi_\beta \phi_\beta^* +
{w_0 \over 2} \phi_\alpha \phi_\alpha \phi_\beta^* \phi_\beta^* \,\Bigr{]}
\end{equation}
with $\phi_{\alpha}$ a complex vector order parameter field.
Summation of repeated indices $\alpha,\,\beta = 1,\,\dots N$ is implied
with $N = 3$ for the Heisenberg case. The deviation from the transition
temperature is given by $r_0^2$ and there are two coupling constants
$u_0$ and $w_0$, both positive for a non--collinear ground state,
that is the two vectors forming the complex vector $\phi$ should
not be parallel. For the non frustrated case there is only one field
and one coupling constant, but for the superfluid $^3$He the above complexity
also arises \cite{Bailin,Jones}. 

\begin{figure}

\centerline{
\psfig{figure=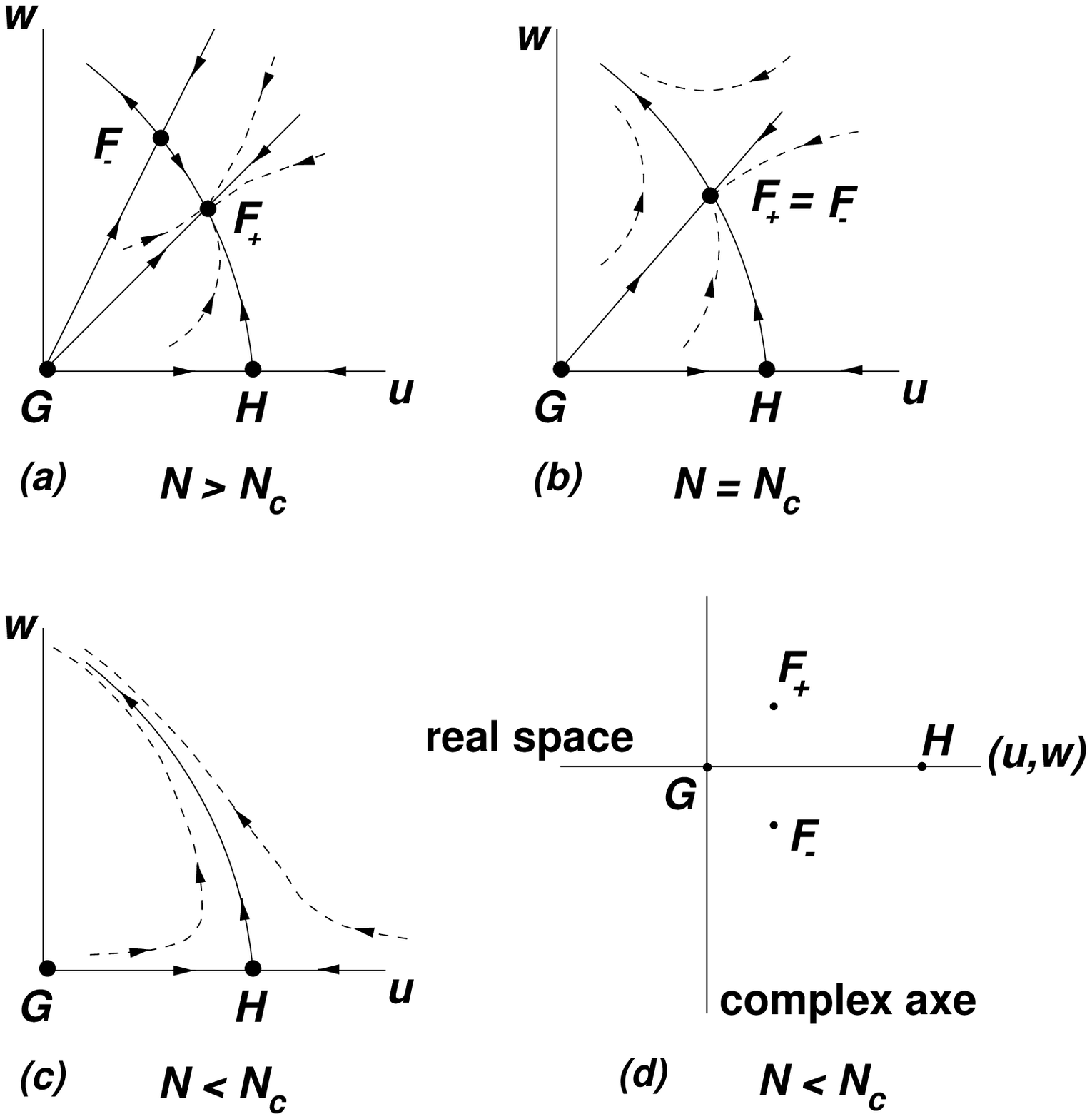,width=6.5cm} }
\caption{\label{figure21}
a), b), c) Hamiltonian flow induced by
renormalization group transformations. The
arrows indicate the direction of flow under iterations.
d) For $N<N_c$ the fixed points $F_+$ and $F_-$ become complex.
}

\vspace{0.9cm}

\centerline{
\psfig{figure=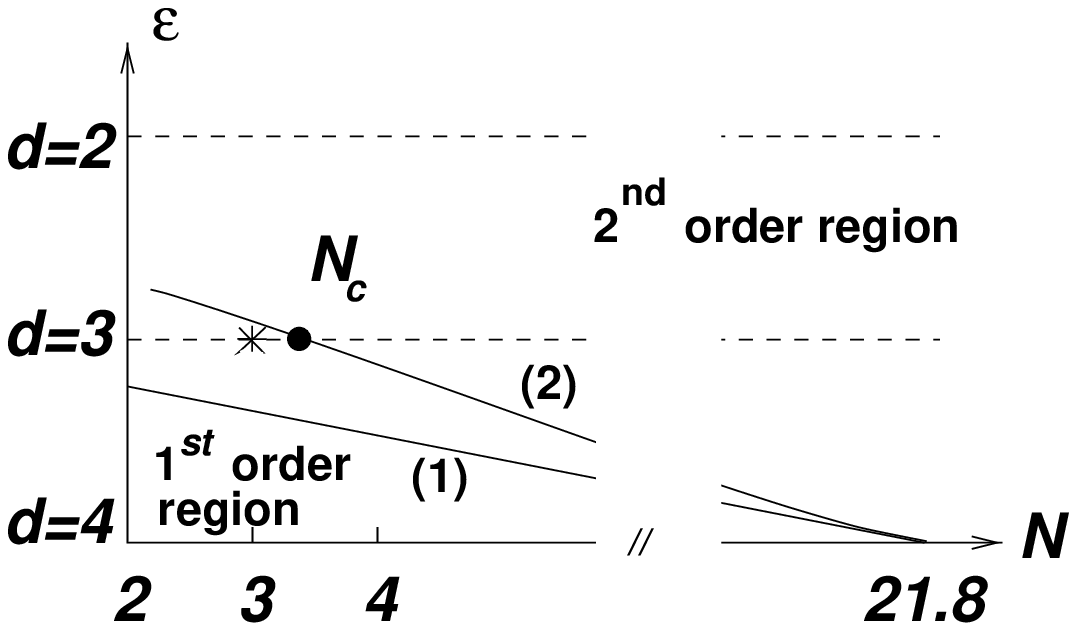,width=6.5cm} }
\caption{\label{figure22}
The critical curve $N_c(d)$ separating the first order region
at small $N$ and large $d$ from the second order region for
large $N$ and low $d$. Curve (1) is the result of first order in $\epsilon
= d - 4$, and curve (2) in second order. The STA at $d=N=3$ is marked
by a cross. In contrast to curve (1) curve (2) lets the physical
relevant systems $N=2,\,3$ stay inside the first order region.
}

\end{figure}

Renormalization group calculations for $d = 4 - \epsilon$ 
\cite{Bailin,Garel,Kawa88} or also directly
for $d = 3$ \cite{Antonenko 94} show that four fixed points exist:
\begin{itemize}
\item[1.] The Gaussian fixed G point at $u^* = v^* = 0$ with mean field
critical exponents.
\item[2.] The $O(2N)$ fixed point H at $w^* = 0$ and $u^* = u_H\ne 0$
with $O(2N)$ exponents (see table \ref{table2}).
\item[3.]Two fixed points $F_+$ and $F_-$ at location ($u_{F_+},w_{F_+})$
and $(u_{F_-},w_{F_-}$) different from zero.
These are the fixed points associated with a new "chiral" universality
class.
\end{itemize}
The existence and stability of the fixed points depend on the number
of components $N$:
\begin{itemize}
\item[a.]  $N>N_c$: four fixed points are present but
three are unstable ($G$, $H$, $F_-$) and the stable one is $F_+$.
Therefore the transition belongs to a new universality class
different from the standard $O(N)$ class.
If the initial point for the RG flow is to the left of the line
($G,\,F_-$), see fig.~\ref{figure21}\thinspace a.
the flow is unstable and the transition will be of first order.
\item[b.] $N=N_c$: the fixed points $F_-$ and $F_+$ coalesce to a 
marginally stable fixed point.
One would think that the transition is ``tricritical'' but the 
exponents are different and not given by the tricritical mean field values.
The reason is that there are two quartic coupling constants not zero,
see Fig.~\ref{figure21}\thinspace b, in contrast to the tricritical
$O(N)$ point where the quartic term disappears and a sextic term takes over
\cite{tricritical}.
\item[c.] $N<N_c$: F$_-$ and $F_+$ move into the complex parameter space,
see fig.~\ref{figure21}\thinspace c and d. The absence of stable fixed
points is interpreted as a signature of a first order transition.
\end{itemize}
The difficulty is to find a reliable value for $N_c$. The dependence on $d$
has been calculated to first order in $\epsilon = 4 - d$ 
\cite{Bailin,Garel} as
\begin{equation}
N_c = \,21.8 - 23.4 \,\epsilon \ .
\end{equation}
In repeating the analysis for the triangular antiferromagnet and choosing
$\epsilon=1$ for $d=3$ Kawamura \cite{Kawa88}
has argued that the $XY$ and Heisenberg cases should be above $N_c$,
motivated by Monte Carlo calculations which supported a second order
transition of a new universality class 
\cite{Kawa92,PlumerHei,LoisonHei,Bhattacharya,Diep89,Loisonbct}
with critical exponents for the Heisenberg case
$\nu=0.59$, $\beta=0.29$ (see table \ref{table1}) 
different from the standard ones.

Of course the results linear or higher in $\epsilon$ are at best
asymptotic \cite{LeGuillou}. Antonenko, Sokolov and Varnashev
extended the analysis to $\epsilon^2$ \cite{Antonenko2}.
For an analysis directly in three dimensions see \cite{Antonenko 94}.
For the critical number of components after resummation 
\begin{equation}
\label{sokolov1}
N_c=3.91(1) \ 
\end{equation}
is obtained by the direct method \cite{Antonenko 94} and $N_c=3.39$ with the 
$\epsilon$ expansion to second order \cite{Antonenko2}. 
In Fig.~\ref{figure22} the results of the RG studies are depicted.
There is a line which separates a region of first order transition
for high dimensions and low $N$ (Fig.~\ref{figure21}c)
from a region of second order transition for low dimensions and high $N$
(Fig.~\ref{figure21}a). The transition on this line is special
(Fig.~\ref{figure21}b) but is not the standard tricritical.
 
Similar considerations have been applied to the
normal to superconducting type II transition. Starting point is
the Ginzburg--Landau Hamiltonian with also two coupling constants,
a quartic coupling and the coupling to the magnetic field.
Expanding to first order in $\epsilon = 4 -d$
Halperin et al. \cite{Halperin} and Chen et al. \cite{Chen} 
arrived at results almost identical to the ones for frustrated magnets.
There are also four fixed points with one stable one
if the number of order parameters $N$ exceeds $N_c =365.9$ for
a fictitious superconductor in 4 dimensions. The unstable $0(2N)$ fixed
point is replaced by an $XY$ fixed point. For $N < N_c$ only
the unstable Gaussian and this unstable $XY$ fixed point are present,
and the question whether the physical relevant $N = 2$ case is
of second order or not can also not be answered, although the value of
$N_c$ in two loop order has been calculated \cite{Tessmann}:
$N_c = 365.90 - 640.76\,\epsilon + O(\epsilon^2)\ $.
If we put $\epsilon=1$ we obtain $N_c(d=3)\approx-275$ which is less
than two and the transition should be of second order.
The nematic-to-smectic-A  transition in
liquid crystals is described by a model similar to the
Ginzburg--Landau Hamiltonian  transition
\cite{De Gennes,Halperin2} and again the problem of second or
first order transition arises \cite{Halperin,Patton}.

The superfluid phase transitions of He$^3$ and the phase transitions
of helimagnets were first studied
with the field theoretical approach (\ref{Hamiltonian})
by Love, More, Jones and Bailin~\cite{Bailin,Jones}
and by Garel and Pfeuty \cite{Garel} respectively.
In principle all the considerations should also be applicable
to these systems, however for He$^3$ the critical region is not really
accessible. Returning to the
frustrated antiferromagnets and accepting the value $N_c=3.91$ for
$d=3$ of eq.~\ref{sokolov1}, the $XY$ and Heisenberg systems
and also the helimagnets should have a first order transition
in three dimensions.

\subsection {Almost second order transitions and complex fixed points}

The discrepancy between the results of RG studies
which indicate a first order transition and the results of Monte Carlo
study which favor a second order transition (see table \ref{table1})
can be removed by using complex fixed points in the RG analysis.
Indeed, as $N_c$ is very close to $N=3$ the fixed points $F_+$ and $F_-$
will not be far off from the real space of parameters $u^*$ and $v^*$
at $N_c$ as depicted in fig.~\ref{figure21}\thinspace d. Zumbach 
\cite{Zumbach93} has studied in detail the influence of a complex fixed 
point on the RG analysis.
The basin of attraction of such a complex fixed point will mimic a
second order transition with modified scaling relations
\begin{eqnarray}
\label{Zumbach2}
2\,\beta &=& \nu\,(d-2+\eta-c_{\beta})\\
\label{Zumbach3}
\gamma &=& \nu\,(2-\eta+c_{\gamma})\ .
\end{eqnarray}
The constants $c_{\beta}$ and $c_{\gamma}$ are corrections of the scaling
relations absent if the fixed point is real. Using these relations to
determine $\eta$ without corrections 
we get a negative value for $\eta$. Since $\eta$ must be
positive \cite{Patashinskii} 
or at least zero we have an estimate for $c_{\beta,\gamma}$
and an indication that the transition is not a real second order one.
We have used this criterion before for $x$--$y$ case \cite{LoisonSchotteXY}.
The essential point is the smallness of $\eta$,
approximately  $\eta \approx 0.03$
for magnets, so that the correction terms $c_{\beta,\gamma}$ are
clearly visible.

A negative $\eta$ is unphysical since unitarity would be violated
in the corresponding quantum field theory \cite{Patashinskii,Zinn89,Schakel}.
For spin glasses
this restriction no longer holds and indeed a negative $\eta$ can be
obtained \cite{Harris76}. Also for superconductors a negative $\eta$
have been found \cite{Chen} influenced by the choice of gauge \cite{Vasil83}.
The critical exponents must be gauge-independent and in a recent study
\cite{Nogueira} it has been shown that $\eta$ is positive for
superconductors.

Using the local potential approximation for the RG the
critical exponent $\nu$ can be calculated for real and complex fixed points
(more precisely for a minimum in the flow)
following Zumbach \cite{Zumbach93}.
In this approximation $\eta$ is zero and
the result will have errors of a few percents, for
example 8\% for the Heisenberg ferromagnet \cite{Zumbach2}.
For the fixed point $F_+$
of the frustrated systems Zumbach obtains $\nu\simeq0.63$ for $N=3$.
This result is compatible
with $\nu=0.59(1)$ obtained by MC calculations for Heisenberg spins on
a stacked triangular lattice (see table \ref{table1}).

By the same approximation Zumbach estimated the interval where
the systems should be under the influence of a complex fixed point as
\begin{equation}
2.58 < N <4.7 \ \>.
\end{equation}
Outside of these limits for $N<2.58$ the transition is of first order
and for $N>4.7$ the transition belongs to the new universality class.
The two values $N_c = 4.7$ for the local potential approximation
and $N_c=3.91(1)$ found by Antonenko and Sokolov \cite{Antonenko 94}
by field theoretical means are not so different.
The minimum limit 2.58 is in agreement with our simulations for
the Stiefel model. The transition is indeed of first order for
$N=2$ \cite{LoisonSchotteXY} and it will be shown here that the
$N=3$ case has a pseudo second order behavior.

\section {Models and simulations}
In this section we introduce briefly different models that we 
use in this work.  A more complete presentation can be found in 
\cite{LoisonSchotteXY}.

\subsection {The STAR model}
\begin{figure}

\centerline{
\psfig{figure=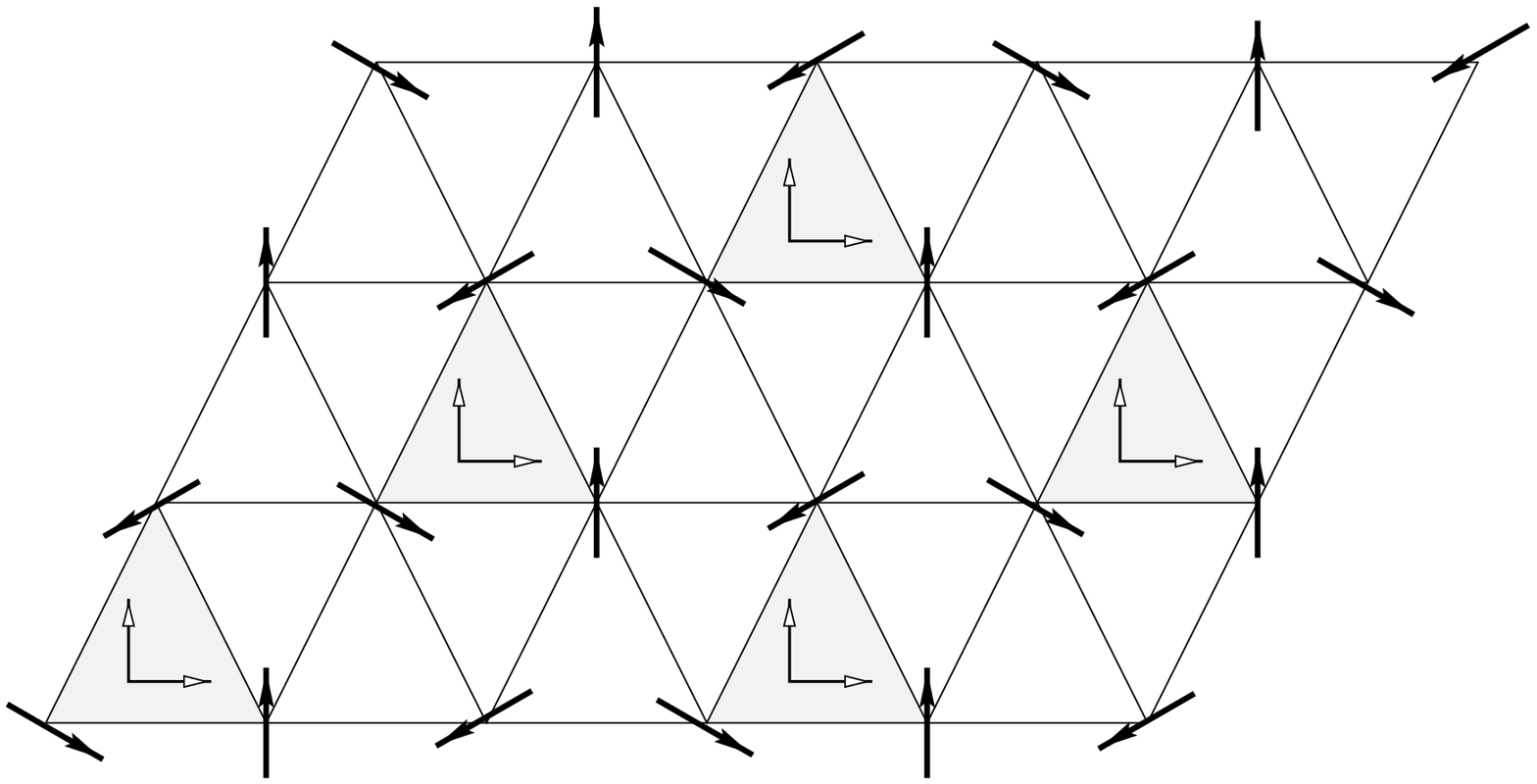,width=6.5cm} }
\vskip 1cm
\caption{\label{figure24}
Ground state configuration for the STA and the STAR model.
The supertriangles are shaded.
The dihedral are drawn at the center of each supertriangle.
}

%


\centerline{
\psfig{figure=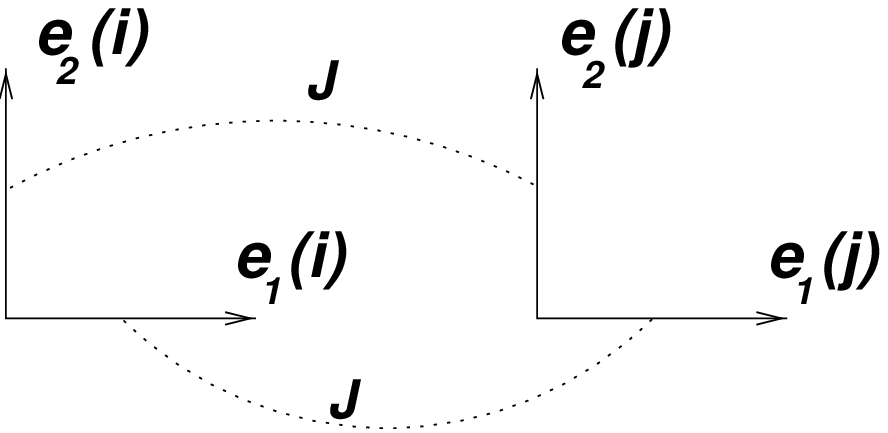,width=6.5cm,height=3.2cm} }
\vskip 0.5cm
\caption{\label{figure26}
The interaction between the ``spins''
or dihedrals $i$ and $j$ of Stiefel's model  $V_{3,2}:$ where
${\bf e}_1(i)$ interacts only with
${\bf e}_1(j)$ and ${\bf e}_2(i)$ with ${\bf e}_2(j)$.
}


\end{figure}

For the stacked triangular antiferromagnet (STA) we take the simplest
Hamiltonian with one exchange interaction constant $J>0$ (antiferromagnetic).
\begin{equation}
\label{tata1}
        H = \sum_{(ij)} J_{ij}\,{\bf S}_{i}.{\bf S}_{j} \ ,
\end{equation}
where ${\bf S}_i$ is a three component spin vector of unit length,
and the sum is over all neighbor spin pairs of the lattice. There are
six nearest neighbor spins in the plane and two in adjacent planes. In
the ground state the spins are in a planar arrangement with the three
spins at the corners of each triangle forming a 120$^{\circ}$ structure
(see Fig.~\ref{figure24}).

For one triangle the spin vectors obey the equation
\begin{equation}
        {\bf S}_{1}+{\bf S}_{2}+{\bf S}_{3} = {\bf 0} \ .
\label{star2}
\end{equation}
where the indices refer to the three corners. To get the STAR model one imposes
this equation as a restriction for the spin directions valid at all
temperatures.
In this theory the local fluctuations violating this constraint
become modes with a gap. Thus they do not contribute to the critical
behavior and can be neglected.
However, one can do this only for selected triangles, for instance the shaded
ones in fig.~\ref{figure24},
otherwise the spin configuration would be completely frozen.

The lattice is partitioned into interacting triangles which do not have common
corners, so that each spin belongs only to one triangle. For the Monte Carlo
simulation one spin direction (two degrees of freedom) is chosen and then the
direction of the second spin selected in the cone of 120$^{\circ}$ around the
first one. The third is determined by the constraint (\ref{star2}). 
All the spins of the supertriangles (the shaded ones) have,
only in the ground state, 
the same orientation. At finite temperatures there is a perfect order within
a supertriangle, but fluctuations of the orientation between these triangles
occur.

The Monte Carlo updating for the state of the supertriangle is done
the following way. First two orthogonal unit vectors are chosen at random
in three dimensions. One needs three Euler angles to do this, the first
$\theta_0$ must be chosen with probability $\sin(\theta_0)\, d \theta_0$
and the other two $\theta_{1,2}$ with probability $\theta_{1,2}$ so that
\begin{eqnarray}
\label{euler1}
{\bf u}&=&\pmatrix{
\sin(\theta_{0}) \cos(\theta_{1})\cr
\sin(\theta_{0}) \sin(\theta_{1})\cr
 \cos(\theta_{0}) \cr}  \\
\nonumber\\
\label{euler2}
{\bf v}&=&\pmatrix{
-\sin(\theta_1)\cos(\theta_2) -\cos(\theta_0) \cos(\theta_1) \sin(\theta_2) \cr
 \cos(\theta_1)\cos(\theta_2) -\cos(\theta_0) \sin(\theta_1) \sin(\theta_2) \cr
 \sin(\theta_0)\sin(\theta_2) \cr} 
\end{eqnarray}
Then the three spin vectors on each supertriangle are determined by
\begin{eqnarray}
{\bf S_1}&=&{\bf u} \nonumber\\
{\bf S_2}&=&\cos(120^{\circ})\, {\bf u} + \sin(120^{\circ})\, {\bf v} \\
{\bf S_3}&=&\cos(240^{\circ})\, {\bf u} + \sin(240^{\circ})\, {\bf v}
\nonumber\ .
\end{eqnarray}

The interaction energy between the spins of this supertriangle with the
spins of the neighboring ones is calculated in the usual way. We follow the
standard Metropolis algorithm to update one supertriangle after the
other. One million MC--steps for equilibration are carried out and
up to six million steps were used for the largest sizes to obtain
reliable averages. Long enough simulations are necessary because the critical
slowing down is strong.

The number of spins is given by $N = L\times L\times L_z$, where $L \times L$
gives the number of spins in one plane and $L_z = 2\,L/3$ the number
of planes. Simulations have been done for $L = 18,\,21,\,24,\,30,\,36,\,42$,
where L must be a multiple of 3 in order to use periodic boundary
conditions and to avoid frustration effects in the planes.

The order parameter $M$ used in the calculations is
\begin{equation}
   M = {1 \over N} \,\sum_{i = 1}^3 |\,M_{i}| \ 
\end{equation}
where the magnetization $M_i$ is defined on one of the three sublattices.
This definition generalizes the one used for two collinear sublattices.

\subsection {The Stiefel model}
The Stiefel model $V_{3,2}$, that is a ``Zweibein'' (or dihedral model) in 
three dimensional space is a further abstraction of the constrained three 
spin system
discussed in the previous section. The three spins at the corner of a triangle
can be taken as a planar or degenerate ``Dreibein'' where the third leg is a
linear combination of the other two and can be left out. The energy is given by
\begin{equation}
\label{tata2}
H = J \sum_{ij}  \Big{[} \ {\bf e}_{1}(i)\cdot{\bf e}_{1}(j) \, + \,
{\bf e}_{2}(i)\cdot {\bf e}_{2}(j) \ \Big{]} 
\end{equation}
where the mutual orthogonal three component unit vectors ${\bf e}_1(i)$
and ${\bf e}_2(i)$ at lattice site $i$ interact with the next pair of
vectors at the neighboring sites $j$. The interaction constant is here
negative to favor alignment of the vectors at different sites. Also a
cubic lattice instead of a hexagonal one can be taken (for further details
see \cite{LoisonSchotteXY}).

We use a single Monte-Carlo cluster algorithm
\cite {wolff 89a}. A cluster of connected spins is constructed 
and updated in the standard way. The first site of the cluster $o$ is
chosen at random together with a random reflection {\bf r}.
Then all neighbors $j$ are visited and added to the cluster with probability
\begin{equation}
P = 1 - \exp(\min\{0,E_1 + E_2\})
\end{equation}
where
\begin{equation}
E_{1,2} \, = \, 2\beta\,({\bf r\cdot e}_{1,2}(o)) \, ({\bf r\cdot e}_{1,2}(j))
\end{equation}
and $\beta=J/T$ until the process stops. Finally all spins of the cluster
are ``reflected'' with respect to the plane $\perp$ to $\bf r$, that is
\begin{equation}
{\bf e}_{1,2}^{new} = {\bf e}_{1,2}^{old}
- 2\, ({\bf e}_{1,2}^{old}\cdot{\bf r}) \, {\bf r} \ .
\end{equation}
It has been demonstrated that this cluster method is very efficient in
reducing the critical slowing down for the $O(N)$ ferromagnet \cite{wolff 89d}.
We thought it should be useful in our case since the order is also 
ferromagnetic,
contrary to the original STA model where the spins have 
a 120$^{\circ}$ structure so that the cluster algorithm described
does not work.
However, we did not observe a decreasing of the critical slowing down.
Indeed on general grounds Sokal et al. \cite{Sokal} have argued
that Wolff's cluster algorithm cannot reduce the critical slowing down
unless the manifold for the order parameter is a sphere as for $O(N)$
ferromagnet. 

In each simulation 5 million measurements were made
after enough single cluster updatings of 1 million steps for equilibration.
Cubic systems with linear dimension $L = 15,\,20,\,25,\,30,\,35,\,40$
were simulated. In comparison with the STAR model $L$ should be multiplied
by $\sqrt{3}$ since one site of the Stiefel model represents three spins.
The equivalent sizes would be 25 to 70.

The order parameter $M$ for this model is
\begin{equation}
M = {1 \over 2 N}\, \sum_{i=1}^2 \,\big|\,M_{i}\big|
\end{equation}
where $M_i$ is the total magnetization
given by the sum of the vectors ${\bf e}_i$ over all sites and $N = L^3$
is the total number.

\subsection {The right handed trihedral model}
If one adds a third vector given by the vector product
\begin{equation}
{\bf e}_{3}(i)={\bf e}_{1}(i) \times {\bf e}_{2}(i) \ .
\end{equation}
and takes an energy formally identical to the previous one
\begin{equation}
H \> = \> J \, \sum_{(ij)} \,\bigl[\, {\bf e}_1(i) \cdot {\bf e}_1(j) \> + \>
{\bf e}_2(i) \cdot {\bf e}_2(j) \> + \>
{\bf e}_3(i) \cdot {\bf e}_3(j) \,\bigr]
\end{equation}
has one constructed a really different model? Since introducing this
new vector ${\bf e}_3$ no degree of freedom has been added
and therefore this system should belong to the same universality class
as the original $V_{3,2}$ Stiefel model. It is not the $V_{3,3}$ Stiefel
model since only right handed systems are constructed. An
additional Ising symmetry is absent \cite{Kunz}
contrary to  the $x$--$y$ or planar case where the ``chirality'' 
as an additional Ising variable is present.

In the work of Azaria et al.  \cite{Delamotte,Delamotte2} 
it has been shown that the interaction
between the third vector, which is absent in the original STA model, will be
automatically generated by the renormalization group in $d = 2 + \epsilon$
expansion and, at the fixed point, should be equal to the first two.

This system has been studied by Loison and Diep \cite{LoisonTriads} before. 
Here we will confirm that the transition is strongly first order. 
The Monte Carlo procedure is similar to the one used for the STAR model. 
In addition to the two orthogonal vectors {\bf u} and {\bf v} of
eq.~\ref{euler1}-\ref{euler2} a third vector given by the vector product
${\bf w} = {\bf u} \times {\bf v}$ is constructed.
The interaction energy between the spins of this trihedral with the
spins of the neighboring trihedral is calculated in the usual way and
we follow the standard Metropolis algorithm to update one trihedral
after the other.
In each simulation 10\thinspace 000 Monte Carlo steps were made for
equilibration and for taking the averages. Cubic system of linear dimension
up to $L = 30$ were simulated. The order parameter $M$ is similar to the
previous case
\begin{equation}
M = {1 \over 3 N}\, \sum_{i=1}^3 \,\big|\,M_{i}\big|
\end{equation}
but one has to take the average of three $M_i$ instead of two.

\section{Numerical method}

\subsection{Finite size scaling for second order transitions}

We use the histogram MC technique
developed by Ferrenberg and Swendsen \cite{Ferren88}
and divide the energy range into 30,000 intervals, 
for more details see \cite{LoisonSchotteXY}.
The errors are determined with the help of the Jackknife procedure 
\cite{Jackknife}.

For each temperature $T$ we calculate the internal energy per site
$\bar E = \langle E \rangle/N$ and the specific heat
$ C = (\langle E^2 \rangle - \langle E \rangle^2) /(N\,T^2)$, where
$\langle \dots \rangle$ indicate the average and $N$ is the number of sites.
Similarly we determine the averages of the order parameter or staggered
magnetization $\bar M = \langle M \rangle$ and the corresponding
susceptibility
$\chi = N\,\bigl(\langle M^2 \rangle - \langle M \rangle^2\bigr)/T$.
The quantities needed besides $\bar M$ for the finite size analysis
are defined below
\begin{eqnarray}
\chi_{2}\, &=&\,\frac{N}{T}\> \langle M^{2} \rangle \\
\chi_{4}\, &=&\,\frac {N^3}{T^3} \> \bigl( \langle M^{4} \rangle \, - \,
	3\,\langle M^{2} \rangle^2\bigr) \\
\label {titi1}
U\,&=&\,1\, - \,
\frac{ \langle M^{4} \rangle }{3\,\langle M^{2} \rangle ^{2}} \\
V_{1}\,&=&\,\frac {\langle ME\rangle }{\langle M \rangle}\> - \,
	\langle E\rangle
\end{eqnarray}
where $\chi_{2}$ is the magnetic susceptibility
and $\chi_{4} $ the fourth derivative of the free energy 
with respect to the magnetic field 
in the high temperature region where the order parameter is zero.
The cumulant $V_1$ is used to obtain the critical exponent $\nu$,
and the fourth order cumulant $U$ to determine the critical temperature.

According to the FSS theory \cite {Barber 83,Ferren91} for a second order 
transition the various quantities just defined should scale for a 
sufficiently large system at the critical temperature $T_c$ as
\begin{eqnarray}
\label{totoc}
\chi_{2}\, &=& \,  g_{\chi_{2}}\ L^{\gamma/\nu}\\
\label{totoc2}
\chi_{4}\, &=& \,  g_{\chi_{4}}\ L^{\gamma_4/\nu}\\
\label{totod}
V_{1}\, &=& \, g_{V_{1}}\ L^{1/\nu}\\
\label{totoe}
\bar M\,  &=& \, g_{M}\ L^{-\beta/\nu}
\end{eqnarray}
where $\beta = 1/T$ and $g$ are constants not dependent on size $L$.
We will not use $\chi$ to determine $\gamma/\nu$ but $\chi_2$ (\ref{totoc})
and $\chi_4$ (\ref{totoc2}) using
\begin{equation}
\label{gamma4}
\gamma_4/\nu \, = \, d \, + \, 2\,\gamma/\nu\ ,
\end{equation}
since the errors are smaller.

\begin{figure}

\vskip 1cm
\centerline{
\psfig{figure=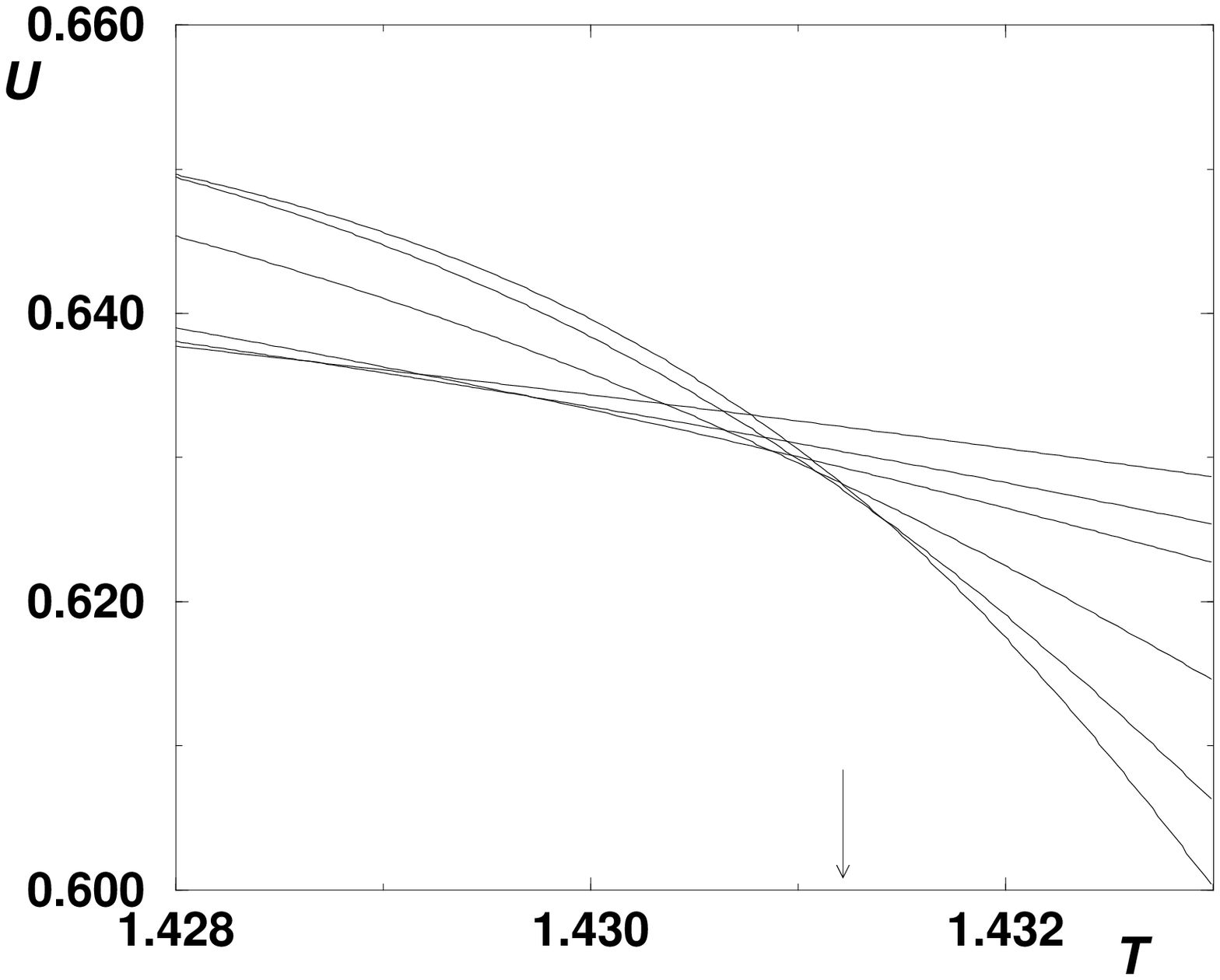,width=6.3cm} }
\caption{\label{figure28}
Binder's parameter $U$ for the STAR model
as function of the temperature for different
sizes $L$ (in the left part of the figure upwards from $L$=18 to $L$=42).
The arrow indicates the critical temperature $T_c$.
}

\vspace{0.8cm}

\centerline{
\psfig{figure=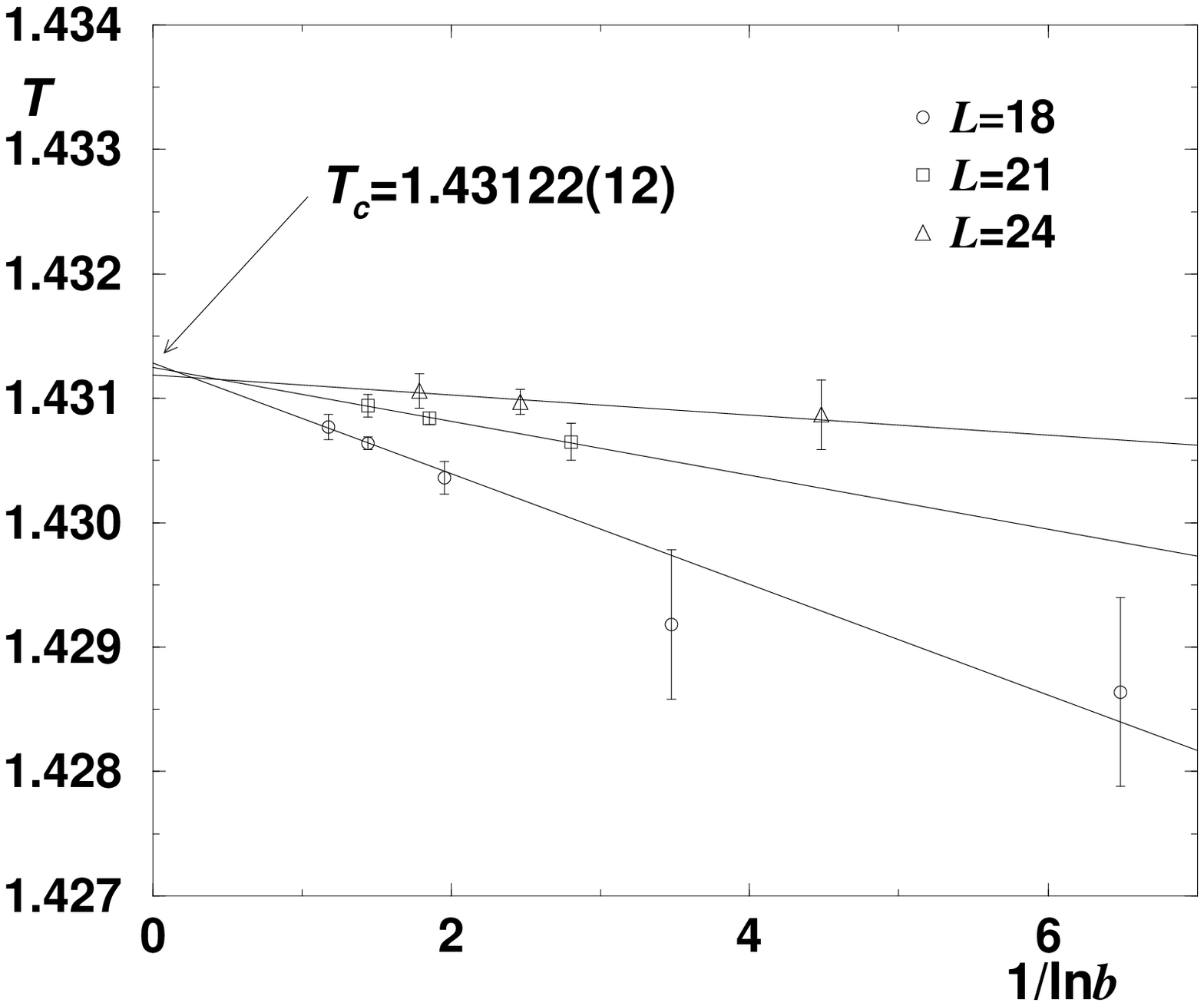,width=6.3cm} }
\caption{\label{figure29}
Estimated $T_c$ for the STAR model
plotted vs inverse logarithm of the scale factor
$b=L'/L$ . For clarity, only the results for $L$=18,\,21,\,24 are shown. The
estimated temperature is $T_c$=1.43122(12).
}

\end{figure}

\begin{figure}

\centerline{
\psfig{figure=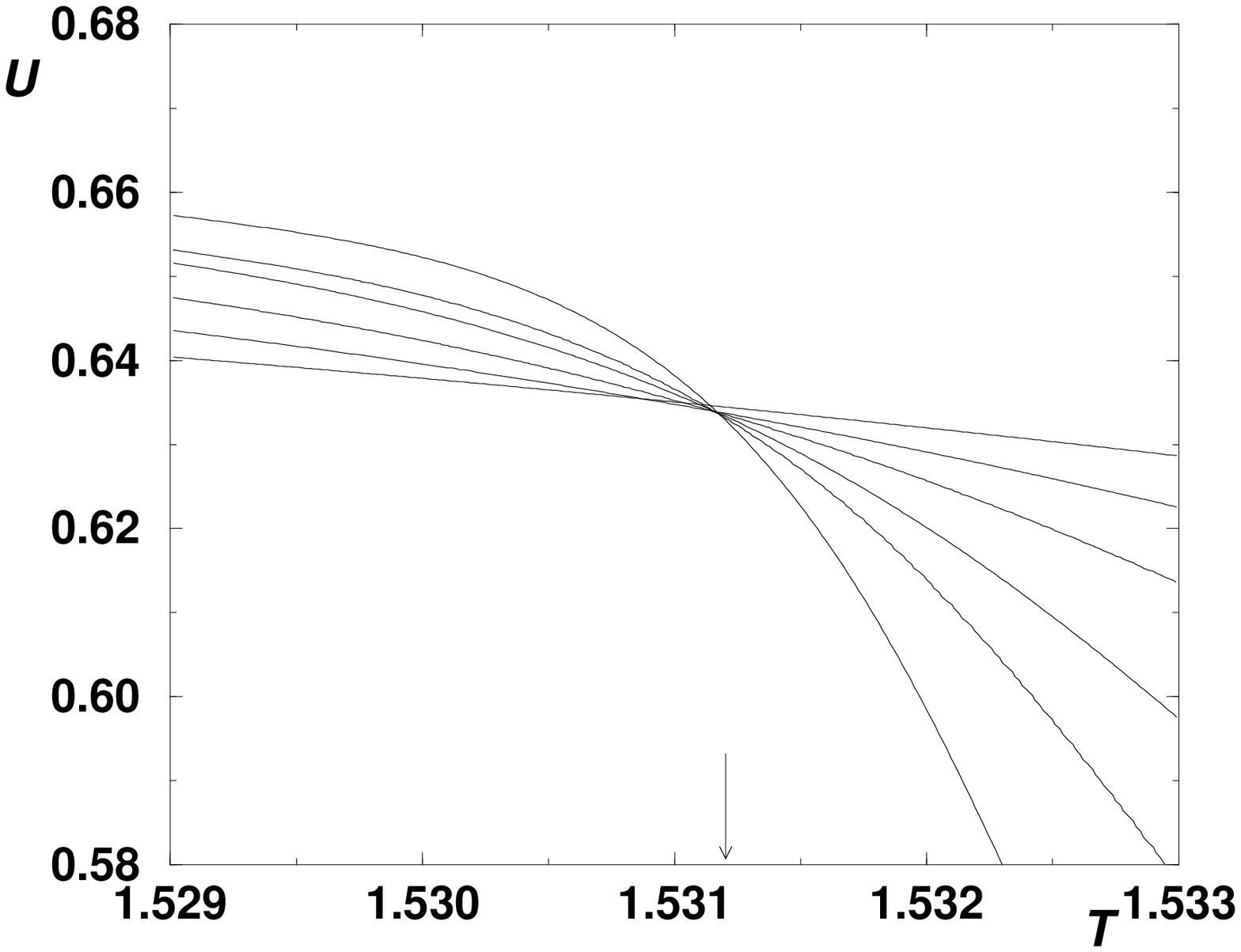,width=6.5cm} }
\caption{\label{figure35}
Binder's parameter $U$ for Stiefel's model $V_{3,2}:$
as function of temperature for different
sizes $L$ (in the left part of the figure upwards from $L=15$ to $L=40$).
The arrow indicates the critical temperature $T_c$.
}
\vspace{1.2cm}

\centerline{
\psfig{figure=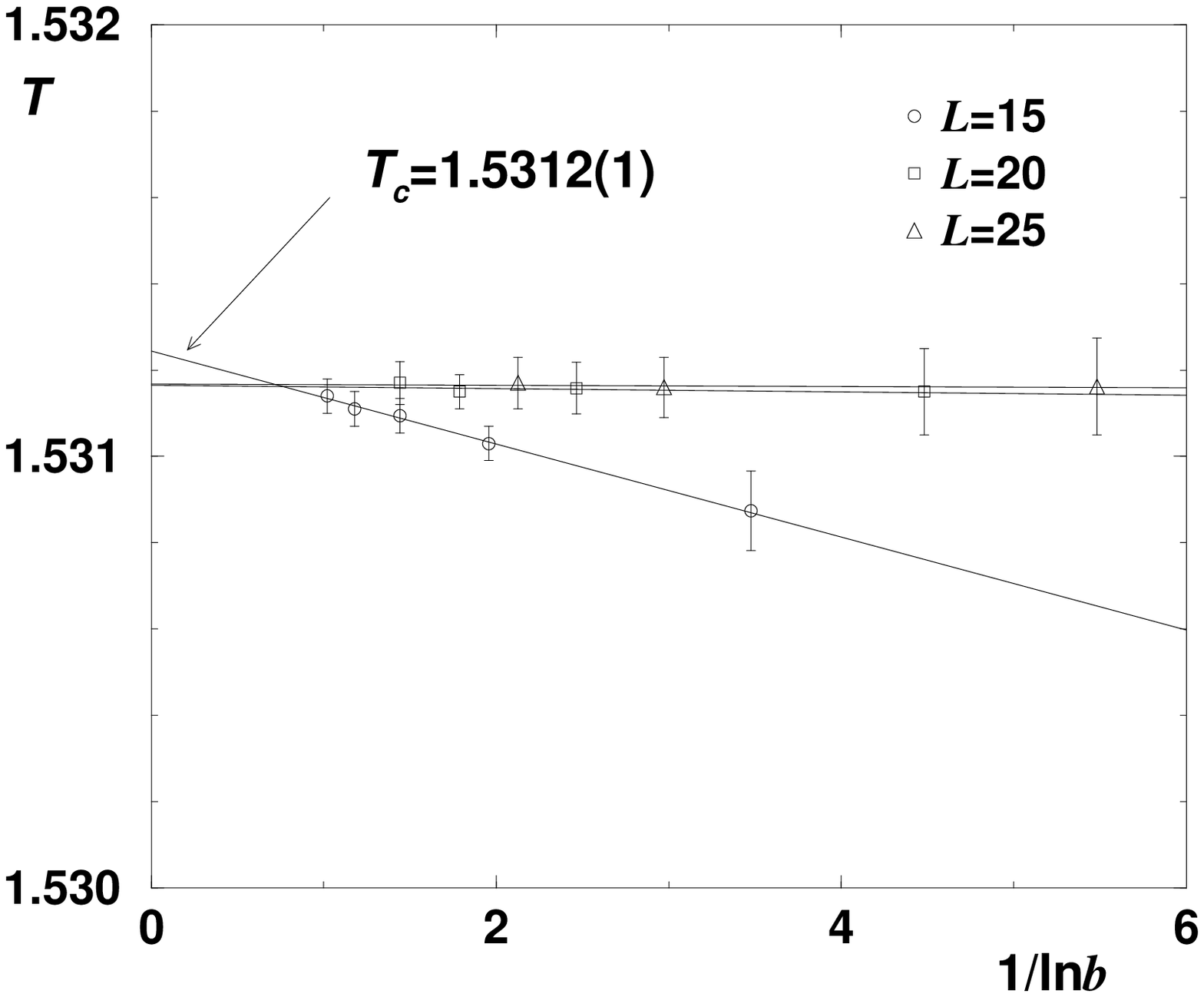,width=6.5cm} }
\caption{\label{figure36}
Estimated $T_c$ for the Stiefel model $V_{3,2}:$
plotted vs inverse logarithm of the scale factor
$b=L'/L\,$. For clarity, only the results for $L=15,\,20,\,25$ are shown. The
estimated temperature is $T_c$=1.5312(1).
}

\end{figure}

To find the critical temperature $T_{c}$ we record the variation of $U$
with $T$ for various system sizes and then locate $T_c$ as the intersection
of these curves \cite{BinderU}, since
the ratio of $U$ for two different lattice sizes $L$
and $L'=bL$ should be 1 at $T_c$, that is
\begin{equation}
\label{BinderU}
{\frac{U_{bL}}{U_{L}}}\Bigg\arrowvert _{T=T_{c}} = \> 1 \> .
\end{equation}
Due to the presence of residual corrections to finite size scaling,
actually one has to extrapolate the results taking the limit 
(ln$b$)$^{-1} \rightarrow  0$ (Fig.~\ref{figure28}, \ref{figure29},
\ref{figure35} and \ref{figure36}).

\subsection{First order transitions}
A first order transition has a different scaling behavior
\cite{Privman,Binder2,Billoire2}.
\begin{itemize}
\item[a.] The histogram $P(E)$ should have a double peak.
\item[b.] Magnetization and energy should show a hysteresis.
\item[c.] The minimum of the fourth order energy cumulant 
$W = 1 - \langle E^4 \rangle /(3\langle E^2 \rangle^2)$ varies as
\begin{equation}
\label{tyty3}
W \, = \, W^* \> + \> b\, L^{-d}
\quad {\rm and} \quad  W^* \neq \mbox{$2\over 3$} \ .
\end{equation}
\end{itemize}
A double peak in P(E) means that at least two states
with different energies coexist at the same temperature.
As a consequence the fourth order energy cumulant cannot be
$2\over 3$. This fact was employed for the smaller sizes in a
preliminary study \cite{LoisonTriads}.
Hysteresis effects have to be expected in the simulation of larger
systems where the two peaks are
well separated since the transition time from one state
to the other grows exponentially with the system size.
For frustrated antiferromagnets we are studying
these criteria are not very helpful. Only for the trihedral model
the hysteresis is clearly visible in
Fig.~\ref{figure37},
indicating a strong first order transition.

\section{Results}     

\subsection{STAR model}
Using (\ref{BinderU}) we first determine $T_c$. $U$ is plotted for different 
sizes from $L = 18$ to $L = 24$ as a function of temperature in 
Fig.~\ref{figure28}.  From the intersections we extrapolate $T_c$ as
\begin{equation}
\label{tgtg}
T_{c}=1.43122(12) \ ,
\end{equation}
see Fig.~\ref{figure29}. The estimate for the universal quantity $U^{*}$ at the
critical temperature is
\begin{equation}
U^*=0.6269(10).
\end{equation}

\begin{figure}

\centerline{
\psfig{figure=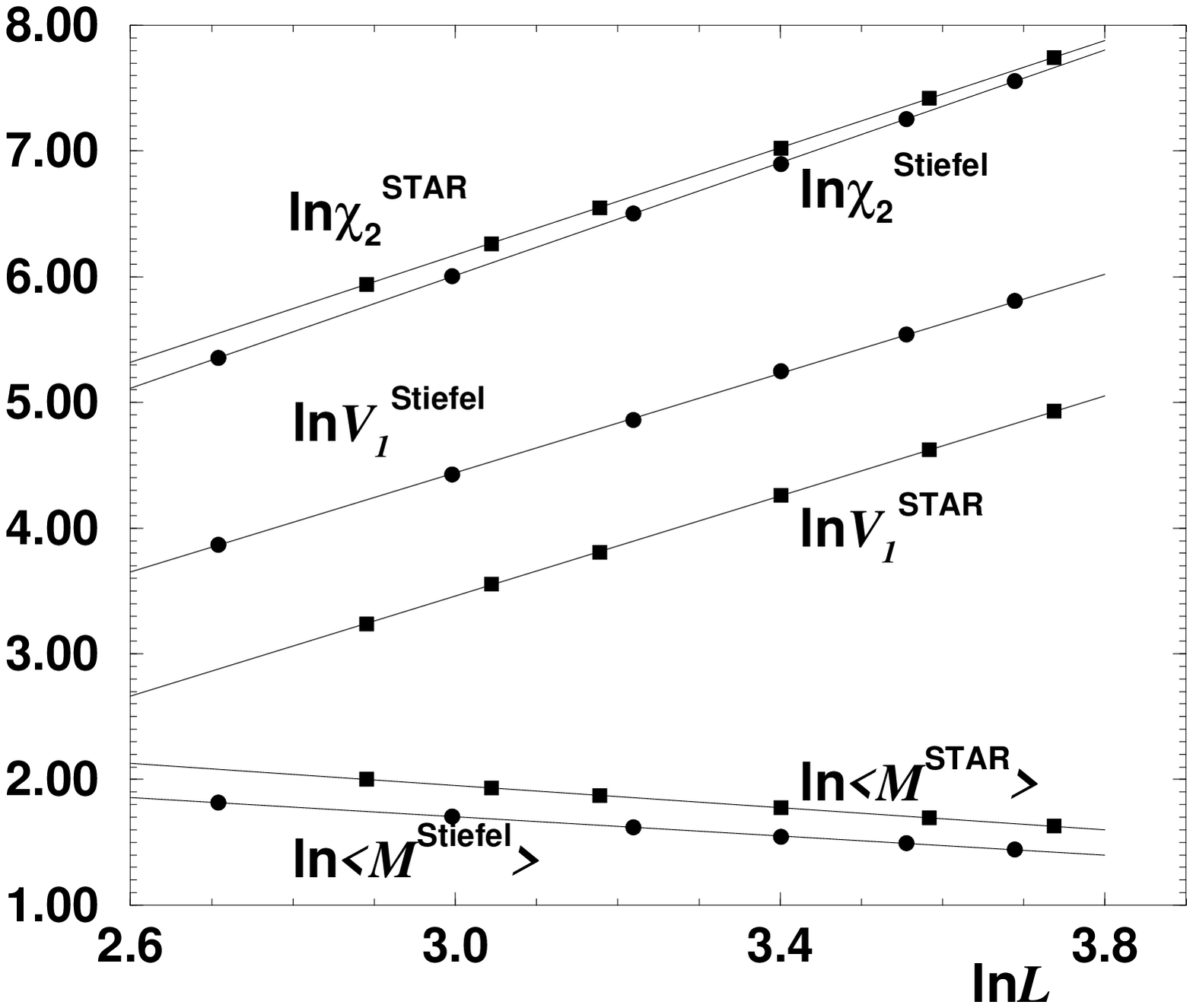,width=6.0cm} }
\caption{\label{figure30}
Values of $\chi_2$, $V_1$ and $\bar M$ as function of $L$
in ln--ln scale at $T_c$ for
the STAR and the Stiefel model $V_{3,2}:$ to get the slopes
$\gamma/\nu$, $1/\nu$ and $\beta/\nu$.
Error bars are smaller than symbols. To obtain a clear figure
values for $\bar M$ are shifted
by +3 units.
}
\vskip 1.3cm

\centerline{
\psfig{figure=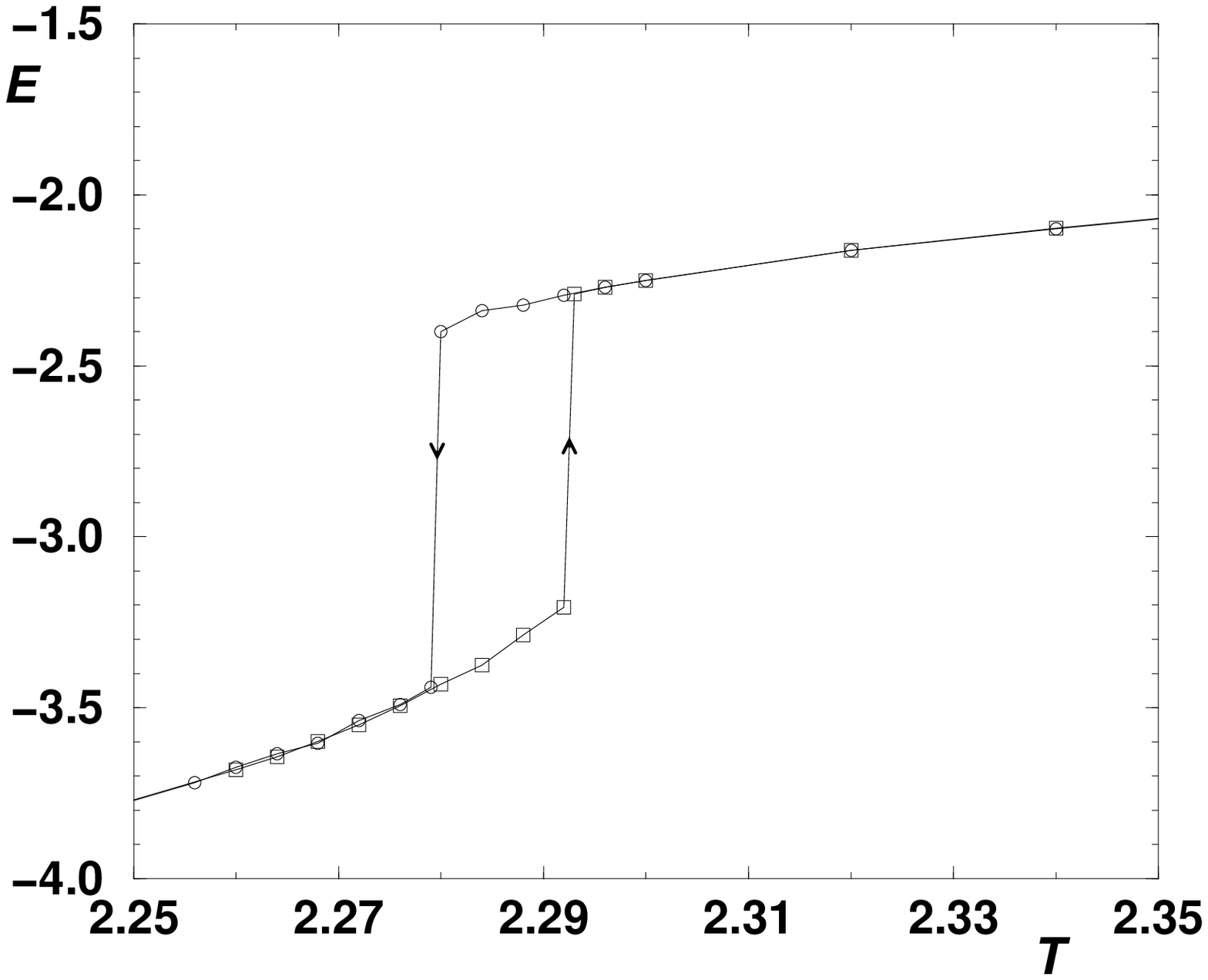,width=6.5cm} }
\caption{\label{figure37}
Hysteresis in the internal energy per spin $E$ versus T for the
right handed trihedral system.
Lines are guides to the eye. The arrows indicate if the MC simulation
is cooling (circle) or heating (square) the system.
}

\end{figure}

With the value of $T_c$ we determine the critical exponents by log--log
fits. We obtain $\nu$ from $V_1$ (Fig.~\ref{figure30}), 
$\gamma/\nu$ and $\gamma_4/\nu$ 
from $\chi_2$ and $\chi_4$ (Fig.~\ref{figure30}, not shown), 
and $\beta/\nu$ from $\bar M$ (Fig.~\ref{figure30}):
\begin{eqnarray}
\nu&=&0.504(10)\\
\gamma/\nu &=& 2.131(13) \\
\label{gamma42}
\gamma_4/\nu &=& 7.252(29)  \\
\beta/\nu &=& 0.439(8)\ .
\end{eqnarray}
The uncertainty of $T_c$ is included in the estimation of the errors.
The value of $\gamma/\nu = 2.126(14)$ found from  $\gamma_4/\nu$ 
using (\ref{gamma4})
is compatible with the one found directly. Combining these results we obtain
$\beta = 0.221(9)$, $\gamma = 1.074(29)$ and from the scaling
relation
\begin{equation}
\label{hyperscaling}
\gamma/\nu = 2 -\eta
\end{equation}
we obtain $\eta=-0.131(13)$. 
The results are summarized in table \ref{table4}.

A negative value of $\eta$ is impossible, for a second order phase
transition it should always be positive \cite{Patashinskii}. 
A slightly negative
value was already discernible in some of the results of the unconstrained
STA--model, see table \ref{table1}. In making use of our previous analysis
of the $x$--$y$ case \cite{LoisonSchotteXY} we conclude that the critical 
behavior should be described by the renormalization flow of a fixed point
in the complex parameter space. The scaling relation for $\eta$ above
is then modified to give a positive value for this critical exponent,
see previous section. Moreover this large negative value cannot come
from only the presence of a complex fixed point. 
Thus we interpret the negative value as
a crossover from the basin of attraction of the complex fixed point
to the true first order transition, see Fig.~\ref{figure39} 

In a previous study of the STAR model, Dobry and Diep \cite{Dobry} obtained
for the exponent $\nu=0.440(20)$
which disagrees with ours 0.504(10). However this seems to be a misprint
since the inverse value is also given and $\nu = 1/2.08 = 0.48$ is within the
statistical errors. However the other exponents seem wrong. 
We repeated these calculations for larger clusters and
with better statistics in order to have evidence that this system shows
a weak first order transition, since unambiguously $\eta$ is really
negative

\subsection{Stiefel's $V_{3,2}$ model}
Following the same procedure as for the STAR model we determine first
$T_c$. The cumulant $U$ plotted as a function of temperature for
different system sizes from $L = 15$ to $L = 40$ is shown in Fig.\ref{figure35}.
The extrapolation, shown in Fig.~\ref{figure36}, for $L = 20$ and $L =25$ 
agrees rather well; the little difference for $L = 15$ indicates that this
lattice size is not yet sufficient for a strictly linear extrapolation.
We obtain
\begin{equation}
\label{tgtg2}
T_{c}=1.5312(1) 
\end{equation}
and for the universal quantity $U$ at $T_c$
\begin{equation}
U^*=0.6326(12).
\end{equation}
In plotting the logarithm of $V_1$, $\chi_2$, $\chi_4$ and
$\bar M$ as function of the logarithm of temperature difference
$(T - T_c)/T_c$, shown in Fig.~\ref{figure30}, 
one finds from the slope
\begin{eqnarray}
\nu&=&0.507(8)\\
\gamma/\nu &=& 2.240(10) \\
\label{gamma43}
\gamma_4/\nu &=& 7.480(22)  \\
\beta/\nu &=& 0.381(6) .
\end{eqnarray}
From $\gamma_4/\nu$ using (\ref{gamma4}) we get $\gamma/\nu = 2.240(11)$
which is the same value found directly from $\chi_2$.
Further we obtain for $\beta = 0.193(4)$ and $\gamma = 1.136(23)$
and with the scaling relation (\ref{hyperscaling}) an even more negative value
$\eta = -0.240(10)$. See for a listing table \ref{table4}, where also
the specific heat exponent $\alpha = 2 - d\nu$ has been added.
The errors given include the uncertainty of the estimate of $T_c$.

The results of Kunz and Zumbach \cite{Kunz} are also given there and are 
compatible with ours. Their $\alpha$ is actually determined by fitting specific
heat data in the high temperature region and $\nu$ with the hyperscaling 
relation. They noticed also that $\eta$ is negative, apparently
their value $-0.10(5)$ is too small, compared to $-0.24(1)$
we found. We cannot share their conclusion that the negative $\eta$
is due to a strong finite size effect not properly taken care of.
We rather take it as an indication that an analysis as a genuine second order
transition leads to inconsistencies and interpret the negative value as
a crossover from the basin of attraction of the complex fixed point
to the true first order transition, see Fig.~\ref{figure39}.

\subsection{Direct trihedral model}

It is known from \cite{LoisonTriads} that the energy distribution $P(E)$ 
near the critical
temperature has a double peak structure. A rough estimate of the
correlation length $\xi_0$ by $1 \over 3$ of the smallest size where
the two peaks are well separated by a region of zero probability gives
$\xi_0 = 6$. Here we determine the energy cumulant at the transition.
Extrapolating $W$ to a system of infinite size we obtain
\begin{equation}
W^{*} \, = \, 0.623(1)
\end{equation}
which of course deviates from the value $2\over 3$ for a second order
transition. It is not difficult to see hysteresis effects. In
Fig.~\ref{figure37} hysteresis for the energy $E$ 
is clearly visible.

Therefore the behavior of a system of trihedrals differs from
the one of dihedrals.
In weakening the interaction between the third components of the trihedrals
one could of course recover the apparent second order nature of the
dihedral system. The first order character of the
transition of pure dihedrals is impossible to show directly.
This system remains always under the influence of the virtual or complex
fixed point for the sizes of systems one can simulate, while the presence
of the third vector allows the system of trihedrals to
stay outside of the neighborhood of this fixed point and the ``true''
first order behavior is seen directly.

\section {Discussion}

\subsection{Summary}

The STAR and the Stiefel model $V_{3,2}$ have slightly
different critical exponents. Looking at table~\ref{table4}, most noticeable 
are the differences for $\beta$ and $\gamma$:
$\beta_{STAR}=0.221(9)$ and $\beta_{Stie}=0.193(4)$,
outside the estimated statistical and systematic errors.
The difference to the values of $\beta \approx 0.286$ for the original
stacked triangular antiferromagnet in table~\ref{table1} is even larger.
Also the non--collinear antiferromagnet on the bct lattice
has a different $\beta$--value (see table 1). Using the scaling relation
$\eta = 2 - d + 2\,\beta/\nu$ the exponent $\eta$ is always negative.
Moreover, a member of the same universality class,
the right handed trihedral model, has a strong first order transition.
This does not correspond to the standard behavior of a universality class
one expects from the seemingly clear evidence for a second order transition
as visible in Fig.~\ref{figure30}
for the STAR and the Stiefel model.

In taking the point of view that the simulations show an extremely
weak first order transition it is very natural to use a field theoretical
description with the usual renormalization group scheme supplemented
by complex fixed points, see Fig.~\ref{figure21}.
Such complex fixed points exist if the number of components $N$ is lower
than a critical one $N_c$, and according to field theoretical analysis
the critical dimension is $N_c \approx 4$. As for a real fixed point
the renormalization flow will be attracted and the system imitates
a second order behavior, however it can escape if the system size $L$
is larger than the largest correlation length $\xi_0$ possible and
then the crossover to first order region is reached.
This crossover phenomenon is known to occur in the two dimensional
Potts model with $q=5$ components \cite{Baxter,Landau}.

In the analysis based on complex fixed points Zumbach \cite{Zumbach93}
showed that the scaling relations should be modified by correction terms
eqs.~(\ref{Zumbach2}-\ref{Zumbach3}). Leaving out these terms the
scaling relations for $\eta$ produces negative values, for
$XY$ spins of the STA model it is slightly negative ($\eta\sim -0.06$
\cite{LoisonSchotteXY}) and for the models studied here the effect is
two or three times larger, see table \ref{table4}.
Another possibility of obtaining  negative values for $\eta$
is to visualize the spin system in the crossover region between a second
to first order transition. Using the same scaling relations as before
the exponent $\eta$ will tend to the value $\eta=-1$ of a ``weak first
order transition'' (table \ref{table2}).
Our MC simulations of the STAR and the Stiefel model are in such a
crossover region otherwise the very negative values for $\eta \approx -0.2$
cannot be understood. The same applies for MC simulations
of helimagnetics on bct lattices \cite{Loisonbct}.

\begin{figure}

\centerline{
\psfig{figure=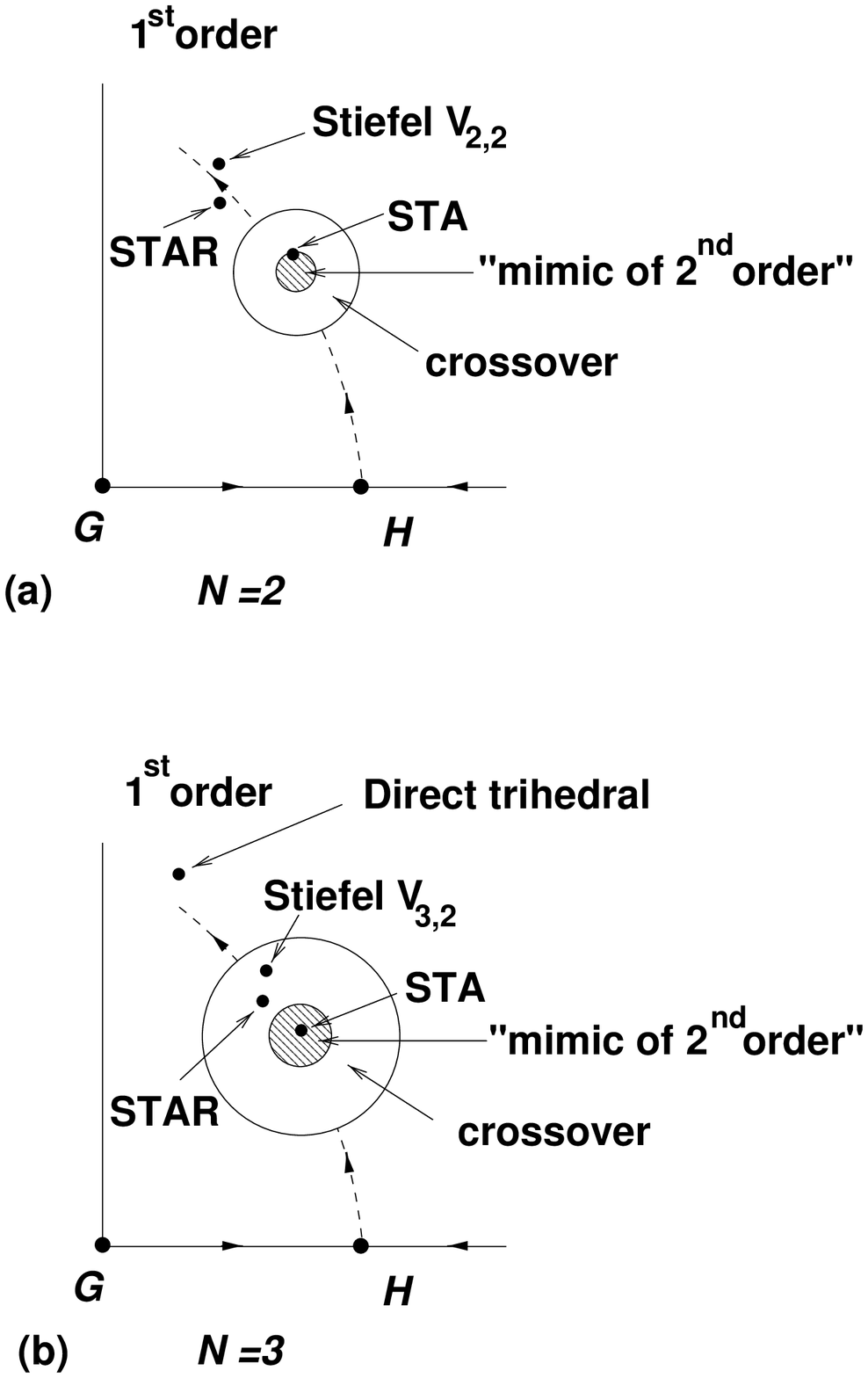,width=6.5cm} }
\vskip 1cm
\caption{\label{figure39}
Hypothesis on the Hamiltonian flow induced by
renormalization--group transformations for $N<N_c$ (see Fig.~\ref{figure21}~c)
showing the smaller influence of the complex
fixed point for the $XY$ spins ($N\!\!=\!2$) compared to the Heisenberg spins
($N\!\!=\!3$).
The arrows give the direction of flow and
the black circles indicate the positions accessible for the different systems
in simulations or experiments. The hatched region is under the influence
of the complex fixed point and the outer circle represents the crossover to
the first order region.
}

\vskip 1.3cm

\centerline{
\psfig{figure=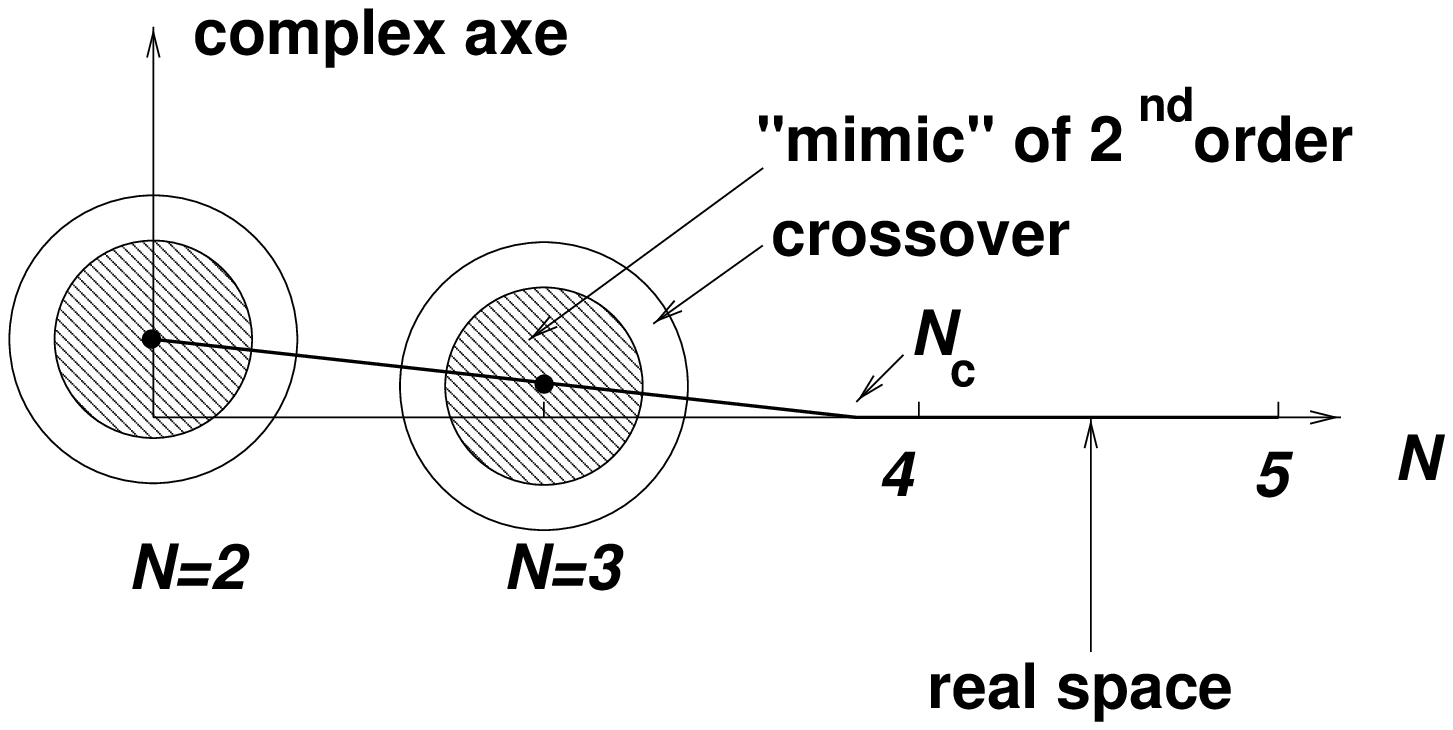,width=6cm} }
\caption{\label{figure41}
The position of the fixed point $F_+$ as function of $N$ for $d=3$ with
its different basins of attraction.
}
\end{figure}

In contrast to the $XY$ case with $N = 2$ the frustrated Heisenberg case is
closer to the critical dimension. That is, the STAR  and the
Stiefel model appear at first sight as second order transitions
whereas in the $XY$ case they show really first order behavior
\cite{LoisonSchotteXY}. The stacked triangular antiferromagnet STA
follows the same trend with a small negative $\eta$ in the $XY$
case and an only slightly negative one in the Heisenberg case table
\ref{table1}. The smaller sphere of influence or basin of attraction
of the complex fixed point in the $XY$ case compared to the Heisenberg case
is depicted in Fig.~\ref{figure39} and Fig.~\ref{figure41}.

Another interpretation following Kawamura \cite{Kawamura98}
takes the diagram in Fig.~\ref{figure21}a as basis, where
an attractive $F_+$ and an unstable fixed point $F_-$ are neighbors.
The initial points in the RG flow for the STAR and the Stiefel model
are placed outside the domain of attraction of the
fixed point $F_+$, that is, to the left of the line ($G,F_-$) in
Fig.~\ref{figure21}a while the STA model should be to the right of this line.
Therefore the STAR and the dihedral model have a first order transition
and the STA model a standard second order transition.
The difficulty is the critical dimension $N_c > 3$ according to
Antonenko and Sokolov \cite{Antonenko 94} and actually
Fig \ref{figure21}c should be the starting point.
Beyond all question is that all the MC simulations give a negative $\eta$
impossible to explain with a real fixed point \cite{Patashinskii}.
One observes also the experimental crossover behavior from second order 
to first order in systems belonging to the $XY$ class
like the holmium, dysprosium or $CsCuCl_3$ (see \cite{LoisonSchotteXY})
which is in disfavor of this interpretation.

\subsection{Experiments}

In real systems because of the omnipresence of planar or axial
anisotropies the behavior of frustrated Heisenberg antiferromagnets
will be difficult to observe. Moreover such systems are usually
quasi one or two dimensional, that is a succession of crossovers will
occur from 2d to 3d and from Heisenberg to Ising or $XY$ behavior. 
To observe the first order behavior the temperature must
be very close to the critical temperature in order to notice
that the correlation length is limited in the basin of attraction
of the complex fixed point $F$. For $XY$ spin systems,
in spite of this limitation the crossover has been observed
in certain materials like Ho \cite{Tindall1}, Dy \cite{Zochowski,Astrom}, 
CuCoCl$_3$ \cite{Weber2} (see also \cite{LoisonSchotteXY}).  

\begin{figure}

\centerline{
\psfig{figure=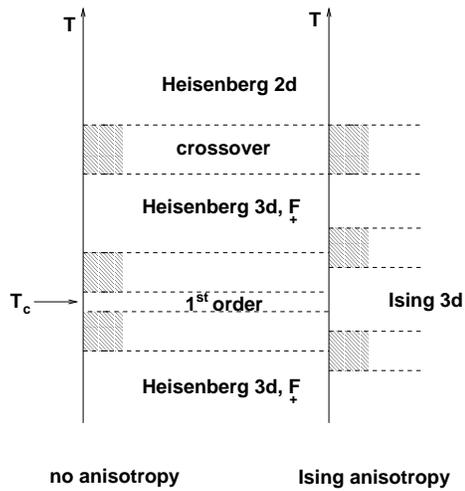,width=6cm} }
\caption{\label{figure40}
Crossover phenomena in experimental systems (not to scale).
There is a crossover to Ising behavior before reaching the first order
region. $F_+$ indicates that the behavior is caused by the complex
fixed point. For real systems it is possible to have
other crossover behaviors due to lattice distortions and
Dzyaloshinsky-Moriya interactions.
}

\end{figure}

Before the first order region is reached in Heisenberg systems
the crossover to Ising or $XY$ behavior prevents a
first order transition of Heisenberg type.
In Fig.~\ref{figure40} a typical scheme is drawn for the crossovers and
transitions of a system with Ising anisotropy, for more details see 
\cite{Diep2} and for other crossovers \cite{Collins1}.

Nevertheless the second order transition can be studied.
A list of results is in Table \ref{table3}:
for VCl$_2$~\cite{Kadowaki}, for VBr$_2$~\cite{Wosnitza}, for
Cu(HCOO)$_2$2CO(ND$_2$)$_2$2D$_2$O~\cite{Koyama}  
and for Fe[S$_2$CN(C$_2$H$_5$)$_2$]$_2$Cl~\cite{DeFotis}. 
For the last two examples the observed exponents might be
influenced by the crossover between 2d to 3d Ising behavior.
The experimental results agree quite well with the MC simulation
in table \ref{table1}.

\subsection {Renormalization group studies in 2+$\epsilon$}

For the STA model studied in $d=2+\epsilon$ expansion using the corresponding
nonlinear $\sigma$--model Azaria et al.\thinspace\cite{Delamotte,Delamotte2}
obtained for all $N$ a stable fixed point $F_{2+\epsilon}$
at a small distance $\epsilon$ from $d=2$.
In comparing results in lowest order of $1/N$ for the $2+\epsilon$
to the $4-\epsilon$ expansion they show that the fixed points found
in the different approaches are equivalent for large $N$.

Specially for $N=3$ a clear evidence by symmetry arguments can be
given that the transition near $d=2$ should be of the standard
ferromagnetic $O(4)$ class. If the transition is of second order
also for $d=3$ it must therefore belong to the $O(4)$ class
with the exponent $\gamma_{O4} = 1.54$ (table \ref{table2})
very different from the value of the STA model $\gamma_{STA} = 1.18$
found by the MC method (table \ref{table1}).
This is not compatible with the Monte Carlo simulations 
and experimental studies (see tables). 

We think that we can rule out the possibility of a 
ferromagnetic $O(4)$ class for $N=3$ if the relevant operators 
are the same for large and low $N$.
The $1/N$ expansion for $\gamma$ (similar arguments hold for all exponents)
for both the ferromagnetic non frustrated (NF)
\cite{Ma} case and frustrated case (F) \cite{Kawa88} are:
\begin{eqnarray}
\gamma_{NF}&=&{2 \over d-2}\,  \Bigl[\, 1-{6 \over N}\, S_d \, \Bigr]\\
\gamma_{F}&=&{2 \over d-2}\,  \Bigl[\, 1-{9 \over N}\, S_d \, \Bigr]
\end{eqnarray}
where
\begin{equation}
S_d={\sin(\pi\,(d-2)/2)\,\Gamma(d-1) \over 2\, \pi\, \Gamma(d/2)^2}
\end{equation}
is a positive quantity.
Thus for $N$ large $\gamma_{NF} > \gamma_{F}$. The reason is the
different breakdown of symmetry from $O(N)$ to $O(N-P)$ with $P=1$ in the
ferromagnetic case and $P=2$ in the frustrated case. When $N$ decreases
the ratio $P/N$ increases also the difference between $\gamma_{NF}$ and
$\gamma_{F}$ will increase and one should have $\gamma_{NF} > \gamma_{F}$
for any $N$ and in particular for $N=3$.

However  it is possible that irrelevant operators in the 
$4-\epsilon$ and $1/N$ expansions become relevant near dimension 
two and low $N$. 
In this case for $N=3$ the transition should be of first order for
dimension three and a little less than three.
Below this limit the transition could be of the "chiral" universality class
and in approaching two dimensions it should become a $O_4$ transition.
This hypothesis has found some support \cite{Delamotte3}. We note that 
at exactly two dimensions, systems are known to be driven by the $O(4)$ 
symmetry, at least at low temperature \cite{Wintel}.

Other RG studies near two dimensions \cite{David,Chubukov} have found more
than a single transition governed by the $O(4)$ fixed point
with an intermediate ``nematic'' phase between the disordered high
temperature region and the ordered $120^{\circ}$ structure at low temperatures. 
This double transition should also occur for the right handed trihedral model
studied here when the interaction between the third components are larger
than between the two others. The ordering of the third vector occurs
before the ordering of the other two vectors and both transitions are
of second order, the first of Heisenberg type and the second
of $XY$ type. The intermediate ``nematic'' phase can also be interpreted
as a ferromagnetic phase and the two transitions characterized as
paramagnetic--collinear and collinear--noncollinear.

Another possibility discussed in the
literature is that the transition is influenced by the presence of topological
defects which are not visible in the continuum and perturbation
theory as in RG ($2+\epsilon$, $4-\epsilon$ and directly in $d=3$)
\cite{Wintel}.
In this work we have assumed that the effects
of topological defects are irrelevant for the critical behavior 
\cite{Kunz,Zumbach95}.

\section {Conclusion}

We have tried to discuss the phase transition of frustrated Heisenberg
spin systems in general terms.
The starting point is Monte Carlo simulations of systems
for which the condition of local rigidity is imposed.
From the finite size analysis we suggest that the transition is of first order.
Our result is that one can rely on the
renormalization group studies and the true behavior is indeed first order.
However one should use the concept of complex fixed points to describe
the almost second order behavior of frustrated Heisenberg spin systems.
In the strict sense there is no ``chiral universality class'' but in
practical terms it exists. We have given reasons why in the usual
Monte Carlo simulations and in experiments this first order transition
is difficult or even impossible to observe.

\section {Acknowledgments}
This work was supported by the Alexander von Humboldt Foundation.
The authors are grateful to Professors B.~Delamotte, G.~Zumbach, H.T.~Diep, 
A.~Dobry, Y.~Holovatch and F.~Nogueira for discussions.
We thank A.I.~Sokolov, E.~Brezin and J. Zinn-Justin 
for the reference of the proof of $\eta \ge 0$, 
S.~Thoms for the reference of the two loop order calculation for 
superconductors and A.~Schakel for the reference
of the gauge dependence of $\eta$ for superconductors. 
Moreover we want to acknowledge L.~Beierlein and M.E.~Myer to a carefully
reading of the manuscript.

\begin{table}[t]
\begin{center}
\begin{tabular}{c|c|c|c|c|c|c}
\hspace{-10pt}
\begin{tabular}{c}system\ \ \end{tabular}
\hspace{-10pt}
&
\begin{tabular}{c}ref.\end{tabular}
\hspace{-10pt}
&
\begin{tabular}{c}$\alpha$\end{tabular}
\hspace{-10pt}
&
\begin{tabular}{c}$\beta$\end{tabular}
\hspace{-10pt}
&
\begin{tabular}{c}$\gamma$\end{tabular}
\hspace{-10pt}
&
\begin{tabular}{c}$\nu$\end{tabular}
\hspace{-10pt}
&
\begin{tabular}{c}$\eta$\end{tabular}
\hspace{-10pt}
\\
\hline
STA&\cite{Kawa92}&0.240(80)&0.300(20)&1.170(70)&0.590(20)&+0.020(180)$^1$
\hspace{-10pt}
\\
\hline
STA&\cite{PlumerHei}&0.242(24)$^2$&0.285(11)&1.185(3)&0.586(8)&-0.033(19)$^1$
\hspace{-10pt}
\\
\hline
STA&\cite{Bhattacharya}&0.245(27)$^2$&0.289(15)&1.176(26)&0.585(9)
&-0.011(14)$^1$
\hspace{-10pt}
\\
\hline
STA&\cite{LoisonHei}&0.230(30)$^2$&0.280(15)&&0.590(10)&0.000(40)$^3$
\hspace{-10pt}
\\
\hline
bct&\cite{Loisonbct}&0.287(30)$^2$&0.247(10)&1.217(32)&0.571(10)&-0.131(18)$^1$
\hspace{-10pt}
\end{tabular}
\end{center}
\caption{\protect\label
{table1}
Critical exponents by Monte Carlo for $O(3)$. STA = stacked triangular  
antiferromagnet, bct = body-centered-tetragonal,
calculated by $^{1)}\;\gamma/\nu=2-\eta\,$, 
$^{2)}\;d \nu=2-\alpha\,$,
$^{3)}\;2\,\beta/\nu=d-2+\eta\,$.
}
\end{table}

\vspace{2cm}

\begin{table}[t]
\begin{center}
\begin{tabular}{c|c|c|c|c|c|c}
\hspace{-10pt}
\begin{tabular}{c}Crystal\end{tabular}
\hspace{-10pt}
&
\begin{tabular}{c}method\end{tabular}
\hspace{-10pt}
&
\begin{tabular}{c}ref.\end{tabular}
\hspace{-10pt}
&
\begin{tabular}{c}$\alpha$\end{tabular}
\hspace{-10pt}
&
\begin{tabular}{c}$\beta$\end{tabular}
\hspace{-10pt}
&
\begin{tabular}{c}$\gamma$\end{tabular}
\hspace{-10pt}
&
\begin{tabular}{c}$\nu$\end{tabular}
\hspace{-10pt}
\\
\hline
VCl$_2$&Neutron&\cite{Kadowaki}&&0.20(2)&1.05(3)&0.62(5)
\hspace{-10pt}
\\
\hline
VBr$_2$&Calorimetry&\cite{Wosnitza}&0.30(5)&&&
\hspace{-10pt}
\\
\hline
CuFUD$^1$&Neutron&\cite{Koyama}&&0.22(2)&&
\hspace{-10pt}
\\
\hline
Insulating$^2$&Neutron&\cite{DeFotis}&&0.24(1)&1.16(3)&
\
\end{tabular}
\end{center}
\caption{\protect\label{table3}
Experimental values of critical exponents for compound VCl$_2$,
VBr$_2$, $^1$Cu(HCOO)$_2$2CO(ND$_2$)$_2$2D$_2$O and 
$^2$Fe[S$_2$CN(C$_2$H$_5$)$_2$]$_2$Cl. 
}
\end{table}

\vspace{2cm}
\begin{table}[t]
\begin{center}
\begin{tabular}{c|c|c|c|c|c}
\hspace{-10pt}
\begin{tabular}{c}symmetry $\ $ \end{tabular}
\hspace{-10pt}
&
\begin{tabular}{c}$\alpha$\end{tabular}
\hspace{-10pt}
&
\begin{tabular}{c}$\beta$\end{tabular}
\hspace{-10pt}
&
\begin{tabular}{c}$\gamma$\end{tabular}
\hspace{-10pt}
&
\begin{tabular}{c}$\nu$\end{tabular}
\hspace{-10pt}
&
\begin{tabular}{c}$\eta$\end{tabular}
\hspace{-10pt}
\\
\hline
$Z_2$&0.107&0.327&1.239&0.631&0.038
\hspace{-10pt}
\\
\hline
$SO(2)$&-0.010&0.348&1.315&0.670&0.039
\hspace{-10pt}
\\
\hline
$SO(3)/SO(2)$&-0.117&0.366&1.386&0.706&0.038
\hspace{-10pt}
\\
\hline
$SO(4)/SO(3)$&-0.213&0.382&1.449&0.738&0.036 
\hspace{-10pt}
\\
\hline
1$st$ order$^1$&1&0&1&1/3&-1$^2$
\hspace{-10pt}
\end{tabular}
\end{center}
\caption{\protect\label{table2}
Critical exponents for the ferromagnetic systems calculated by RG
{\protect \cite{Antonenko1}}. 
$^1$We cannot define exponents in a first order transition, however in the
case of a weak first order transition the exponents found in MC and in 
experiments must tend to these values.
$^2$calculated by $\gamma/\nu=2-\eta$.
}
\end{table}

\vspace{2cm}

\begin{table}[t]
\begin{center}
\begin{tabular}{c|c|c|c|c|c|c|c}
\hspace{-10pt}
\begin{tabular}{c}system\ \ \end{tabular}
\hspace{-10pt}
&
\begin{tabular}{c}ref.\end{tabular}
\hspace{-10pt}
&
\begin{tabular}{c}$\alpha$\end{tabular}
\hspace{-10pt}
&
\begin{tabular}{c}$\beta$\end{tabular}
\hspace{-10pt}
&
\begin{tabular}{c}$\gamma$\end{tabular}
\hspace{-10pt}
&
\begin{tabular}{c}$\nu$\end{tabular}
\hspace{-10pt}
&
\begin{tabular}{c}$\eta$\end{tabular}
\hspace{-10pt}
&
\begin{tabular}{c}$T_c$\end{tabular}
\hspace{-10pt}
\\
\hline
STAR&--&0.488(30)$^2$&0.221(9)&1.074(29)&0.504(10)&-0.131(13)$^1$
&1.43122(12)
\hspace{-10pt}
\\
\hline
$V_{3,2}$&--&0.479(24)$^2$&0.193(4)&1.136(23)&0.507(8)&-0.240(10)$^1$
&1.5312(1)
\hspace{-10pt}
\\
\hline
$V_{3,2}$&\cite{Kunz}&0.460(30)&&1.100(100)&0.515(10)&-0.100(50)&1.532
\hspace{-10pt}
\end{tabular}
\end{center}
\caption{\protect
\label{table4}
Critical exponents by Monte Carlo for the STAR and the Stiefel model 
$V_{3,2}$.
$^1$calculated by $\gamma/\nu=2-\eta$.
$^2$calculated by $d \nu=2-\alpha$.
}
\end{table}

\newpage
\twocolumn
\newpage
\newpage
\newpage
\newpage
\newpage
\newpage

\end{document}